\documentclass[twocolumn,aps,pra]{revtex4}
\usepackage{amsmath,amssymb}
\usepackage{graphicx}
\usepackage[usenames]{color}

\begin{document}

\title{Bose Hubbard Model in a Strong Effective Magnetic Field: Emergence of a Chiral Mott Insulator Ground State}

\author{Arya Dhar$^1$, Maheswar Maji$^2$, Tapan Mishra$^{1,7}$, R. V. Pai$^3$, Subroto Mukerjee$^{2,4}$ and
Arun Paramekanti$^{2,5,6,7}$}
\affiliation{$^1$ Indian Institute of Astrophysics, Bangalore 560 034, India}
\affiliation{$^2$ Department of Physics and CQIQC, 
Indian Institute of Science, Bangalore 560 012, India}
\affiliation{$^3$ Department of Physics, Goa University, Taleigao Plateau, Goa 403 206, India}
\affiliation{$^5$ Department of Physics, University of Toronto, Toronto, Ontario, Canada M5S 1A7}
\affiliation{$^6$ Canadian Institute for Advanced Research, Toronto, Ontario, M5G 1Z8, Canada}
\affiliation{$^7$ International Center for Theoretical Sciences, Bangalore 560 012, India}
\begin{abstract}
Motivated by experiments on Josephson junction arrays, and cold atoms in an optical lattice 
in a synthetic magnetic field, we study the
``fully frustrated'' Bose-Hubbard (FFBH)
model with half a magnetic flux quantum per plaquette.
We obtain the phase diagram of this model on a $2$-leg ladder at integer filling via the density matrix 
renormalization group approach, complemented by Monte Carlo 
simulations on an effective classical XY model. The ground
state at intermediate correlations
is consistently shown to be a chiral Mott insulator (CMI)
with a gap to all excitations and staggered loop
currents which spontaneously break time reversal symmetry.
We characterize the CMI state
as a vortex supersolid or an indirect exciton condensate,
and discuss various experimental implications.
\end{abstract}

\maketitle
The simplest model to understand strongly correlated bosons is the
Bose-Hubbard (BH) model \cite{fisher.prb1989} which describes bosons hopping
on a lattice and interacting
via a local repulsive interaction. With increasing repulsion, at integer filling,
its ground state undergoes
a superfluid to Mott insulator quantum phase transition
which has been studied using ultracold atoms in an optical lattice \cite{greiner.nature2002}.

Remarkably, recent experiments have used two-photon
Raman transitions to
create a uniform or staggered ``synthetic magnetic field'' for neutral atoms 
\cite{spielman.nature2009},
permitting one to access large magnetic fields for lattice bosons.
The multiple degenerate minima in the resulting Hofstadter spectrum
can be populated by
non-interacting bosons in many ways. Repulsive interactions quench
this ``kinetic frustration'', leading to
unconventional
superfluids \cite{hemmerich.naturephys2011,ffxy.theory,lim.2008,sengupta.epl2011},
or quantum Hall liquids \cite{demlerfqhe.prl2005}.
Tuning the sign of the atom hopping amplitude
or populating higher bands also leads to such frustrated bosonic fluids
\cite{hemmerich.naturephys2011}.
These developments motivate us to study
the interplay of {\it strong correlations and frustration} in the fully frustrated Bose-Hubbard (FFBH), 
with half a ``magnetic flux'' quantum per plaquette \cite{ffxy.theory,lim.2008,sengupta.epl2011}.
At large integer filling, the FFBH is also the simplest
quantum variant of the classical fully frustrated XY (FFXY) model
\cite{jayaprakash.prb1983,olsson.prl1995} of Josephson junction arrays (JJAs)
\cite{mooij}.

\begin{figure}[t]
\includegraphics[width=2.8in]{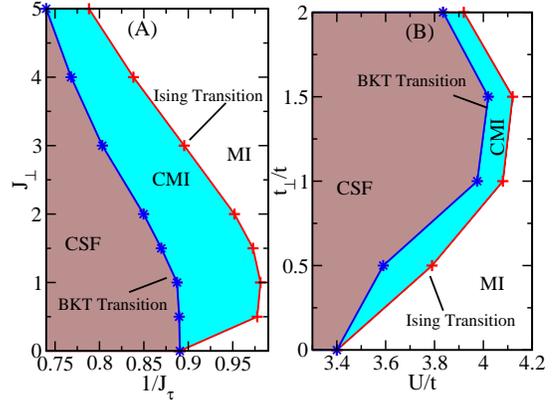}
\caption{(Color online) (A) Phase diagram of the effective classical model $H_{\rm XY}$,
with $J_\tau=J_\parallel$, obtained via
MC simulations (see text for details). (B) Phase diagram of the FFBH model in Eqn.~\ref{LadderHam} obtained using DMRG.
Both models exhibit
a chiral Mott insulator (CMI) state sandwiched between a chiral superfluid (CSF)
and an ordinary Mott insulator (MI). ($1/J_\tau$ in the XY model $\sim \sqrt{U/t}$ in the 
FFBH model.\cite{supp})}
\label{Fig:classical_phased}
\end{figure}

Here, we obtain the phase diagram shown in Fig.~\ref{Fig:classical_phased} 
of the FFBH model at integer filling on a $2$-leg
ladder using the density matrix renormalization group (DMRG)
method \cite{whitedmrg.prl1992} and
Monte Carlo (MC) simulations.
Our key result
is that the ground state of the FFBH and quantum FFXY models
at intermediate Hubbard repulsion
is a {\it chiral Mott Insulator} (CMI).
The CMI  has a nonzero charge gap, and simultaneously
supports staggered loop currents that {\it spontaneously} break time reversal symmetry.
With increasing repulsion, the CMI undergoes an Ising transition into an ordinary Mott insulator (MI)
where the loop currents
vanish. Weakening the repulsion leads to a Berezinskii-Kosterlitz-Thouless (BKT) \cite{kosterlitz.jpc1973}
transition out of the CMI into a previously studied chiral superfluid (CSF) phase \cite{old.csf}
which retains current order. We
show that the CMI may be viewed as a
vortex supersolid or an exciton condensate, 
and discuss
the loop current, the charge gap, and the momentum distribution across the phase diagram.

\begin{figure}
\includegraphics[width=3in,height=1.25in]{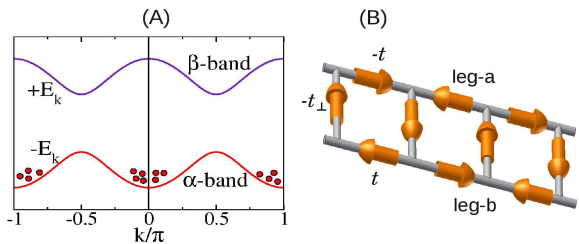}
\caption{(Color online) (A) Dispersion of the FFBH model at $U=0$, 
with two degenerate minima in the low
energy $\alpha$-band.
 Interactions force an equal number of bosons (on average) to condense into each of the two minima.
 (B) Alternating pattern of plaquette currents in the presence of chiral order.}
\label{Fig:dispersion}
\end{figure}

{\it Fully Frustrated Bose-Hubbard Ladder. ---}
The Hamiltonian of the FFBH model on a  2-leg ladder is
\begin{eqnarray}
H \!&=&\!\! -t \sum_x (a^\dagger_x a^{\phantom \dagger}_{x+1} \!+\! a^\dagger_{x+1} a^{\phantom \dagger}_{x})
\!+\! t \sum_x (b^\dagger_x b^{\phantom \dagger}_{x+1} \!+\! b^\dagger_{x+1} b^{\phantom \dagger}_{x}) \nonumber \\
\!&-&\!\! t_\perp
\sum_x (a^\dagger_x b^{\phantom \dagger}_{x} + b^\dagger_{x} a^{\phantom \dagger}_{x}) + \frac{U}{2} \sum_x (n^2_{a,x} + n^2_{b,x}),
\label{LadderHam}
\end{eqnarray}
where $a$ and $b$ label the two legs of the ladder (see Fig.~\ref{Fig:dispersion}), $t_\perp$ couples the two legs, and $U$
is the local boson repulsion. The opposite signs of the hopping amplitude ($\pm t$) on the two legs leads to an Aharonov-Bohm phase of $\pi$ for a boson hopping around an elementary plaquette
\cite{foot1}.

For $U\!=\!0$,
the boson dispersion (in Fig.~\ref{Fig:dispersion} (A))
exhibits two bands with the lowest ($\alpha$) band having degenerate
minima at momenta $k\!=\!0,\pi$.
This leads to a large degeneracy of many-body ground states --- the ground state for $N$ bosons corresponds
to having $N_1$ bosons in one minimum and $(N\!-\!N_1)$ in the other for any $N_1 \leq N$
--- which is broken
by the repulsion.. 
The minimum at $k\!=\!0$ ($k\!=\!\pi$) has a 
wavefunction that mainly resides on leg-$a$ (leg-$b$). Since 
the Hubbard repulsion
favors a uniform density, it
prefers an  {\it equal} number of bosons at $k=0,\pi$.
A mean field Bose condensed state thus takes the form
\begin{equation}
|\psi\rangle = \frac{1}{\sqrt{N!}}\left[{\rm e}^{i \varphi} (\alpha_{0}^\dagger +
{\rm e}^{i\theta} \alpha_{\pi}^\dagger)\right]^N |0\rangle.
\label{Eq:sfwavefn}
\end{equation}
Here $\varphi$ is the $U(1)$ condensate phase, $\theta$ is the relative
phase between the two modes, and
$\alpha^\dagger_{0,\pi}$ creates quasiparticles at $k\!=\! 0,\pi$.

For small $U$, Hartree theory \cite{lim.2008,supp} 
shows $\theta=\pm \pi/2$, while $\varphi$ has (nonuniversal) power law order. 
This Luttinger liquid is the CSF - it supports the long-range
staggered current pattern in Fig.1(B). The
two signs of $\theta$ correspond to patterns related by time-reversal
or unit lattice translation.
For very large $U$, both $\theta$ and $\varphi$ are disordered, leading
to an ordinary MI which respects all the symmetries of $H$.
Remarkably, for intermediate $U$, we find that $\varphi$ is disordered leading to loss
of superfluidity, while $\theta$ is pinned at $\pm \pi/2$,
spontaneously breaking (Ising) time reversal symmetry. This {\it fully gapped} 
intermediate state is the CMI.
This goes beyond mean field theory \cite{lim.2008}
which predicts a direct CSF-MI transition \cite{supp}.

{\it Physical pictures for the CMI. ---}
The CSF, with
staggered currents depicted in Fig.~\ref{Fig:dispersion} (B), is best 
viewed as a vortex crystal where
vortices
and antivortices are nucleated by the presence of frustration, and locked into an
`antiferromagnetic' pattern due to
the intervortex repulsion. At large $U$, this crystal melts and the vortices completely
delocalize - this vortex superfluid is well known to be simply a 
dual description of the ordinary MI \cite{fisherlee.prb1989}.
However if a {\it small} number of defect vortices in
the vortex crystal delocalize and condense, they kill superfluidity but
preserve the background vortex crystallinity.
This {\it vortex supersolid} is the dual description of the CMI.

A different but equivalent picture emerges if we start from the usual MI at
large $U$ which
supports {\it charge gapped} particle and hole excitations (adding or
removing bosons). These excitations have 
degenerate dispersion minima
at $k=0,\pi$ as in Fig.~1(A), similar to the original noninteracting bosons.
Decreasing $U$ decreases the MI charge gap. If the charge gap vanishes, the resulting
gapless particles and holes at $k=0,\pi$ could yield a Bose condensed (or power-law) 
superfluid. However,
a precursor phase emerges from first condensing 
a {\it neutral indirect exciton}, composed of a particle and a hole at different momenta
($k\!=\!0$ and $k\!=\!\pi$), while the particles and holes are still gapped. The CMI is
precisely this intervening `exciton condensate' \cite{supp}.


{\it Effective bilayer XY model. ---} 
To quantitatively flesh out the phase diagram 
described above, 
we first study the FFBH model at large fillings, where it is
equivalent to a quantum FFXY model used to describe
JJAs of charge $2e$ Cooper pairs with an Aharonov-Bohm flux
of $h c / 4 e$ per plaquette. 
The quantum FFXY Hamiltonian in turn maps on to an effective classical model on a
`space-time lattice' leading to a classical 2D bilayer square lattice model \cite{supp}
$
H_{\rm XY} = - \sum_{i, \delta} J_\delta \cos \left( \varphi_i - \varphi_{i+\delta}\right),
\label{Eq:classham}
$
where $\varphi_i$ are the boson phases, and $(i,i\! +\! \delta)$
denote nearest neighbour sites along $\delta$. The couplings $J_\delta$ take on values
$\pm J_\parallel$ on the two legs, $J_\perp$ on the rungs linking the two layers, and $J_\tau$
in the imaginary time
direction \cite{supp}. (We choose the `time step' in the imaginary time direction to set 
$J_\parallel = J_\tau$ \cite{supp}.)
Phase ordering leads to a superfluid, while the fully paramagnetic phase of
$H_{\rm XY}$ is the ordinary MI.
Based on small system studies of $H_{\rm XY}$ \cite{granato.prb1993},
it has been argued that the isotropic case
$J_\perp \! =\! J_\parallel$ exhibits
a single phase transition with novel exponents, while the highly anisotropic case harbors two separate
transitions \cite{granato.prb1993}. Here we use extensive MC simulations, on
$L \! \times\! L \! \times\!  2$
bilayers with $L\!=\! 16$-$64$,
to obtain the phase diagram shown in Fig.~1(A). We
find three phases: the CSF, the regular MI, and an intervening CMI
for a wide range of $J_\perp$ {\it including} the isotropic point $J_\perp\! =\! J_\parallel$.
We show that CSF-CMI and CMI-MI phase transitions are BKT and 2D Ising  transitions respectively.

\begin{figure}
\includegraphics[width=2.8in,height=1.7in]{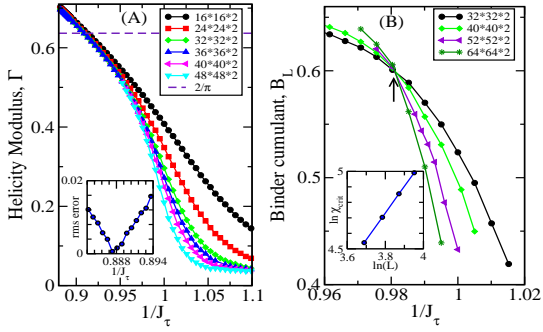}
\caption{(Color online)
(A) Helicity modulus $\Gamma$ versus $1/J_\tau$ for different system sizes
for $J_\perp\!=\!1$. (A-Inset) RMS error of fit to the BKT finite size scaling form of $\Gamma$
shows a deep minimum \cite{supp} at the transition, at $1/J_\tau \!=\! 0.887(1)$, and yields a jump
$\Delta \Gamma\! \approx\! 0.637$,
close to the BKT
value $2/\pi$. (B)
Binder cumulants for the staggered current versus $1/J_\tau$ (for different $L$
for $J_\perp\!=\!1$) intersecting at a continuous transition at $1/J_\tau\! =\! 0.981(4)$.
(B-inset) Critical susceptibility versus $L$ gives the ratio of critical exponents $\gamma/\nu \! \approx\! 1.72$,
very close to 2D Ising value $\gamma/\nu\!=\!7/4$. Error bars are smaller than the symbol sizes. }
\label{Fig:classical_trans}
\end{figure}

Fig.~3 shows the MC data for $J_\perp\! =\! 1$. Similar data was also obtained for 
various $J_\perp/J_\tau$. Fig.~3(A) shows that
the helicity modulus $\Gamma$ (related to the
superfluid density) has an increasingly abrupt change with $1/J_\tau$
for increasing $L$, indicative of a jump as at a BKT transition. If the transition out of the CSF is
indeed a BKT transition,
$\Gamma$  can be fit to the finite size scaling form
$\Gamma (L) \! =\!  A \left( 1\! + \! \frac{1}{2 (\log L + C)} \right)$
(with fit parameters $A$,$C$) right at the transition point,
with $A$ taking on the universal value of $2/\pi$,
while $C$ is a non-universal constant \cite{weber_minnhagen,olsson.prl1995}.
Fitting $\Gamma(L)$ to this form,
we find that the error to this fit shows a sharp
minimum \cite{weber_minnhagen,olsson.prl1995} at a certain $1/J_\tau$
(Fig.\ref{Fig:classical_trans} inset), with $A\! \approx\! 2/\pi$ at this
dip. This not only allows us to precisely locate the transition out of
the CSF state, but also {\it confirms}
its BKT nature.

To check for staggered loop currents,
we compute the Binder cumulant
$B_L = \left(1 - \langle m^4 \rangle_L/3 \langle m^2 \rangle^2_L \right)$, for
the order parameter 
$
m = \frac{1}{L^2}\sum_{i\tau} \left( -1 \right)^i J_{i\tau},
$
where $J_{i,\tau}$ is the current around a spatial plaquette. For small $1/J_\tau$,
we find $B_L \to 2/3$ indicating
long range current order, while $B_L \to 0$ for large $1/J_\tau$ indicating absence
of loop currents. Fig.~\ref{Fig:classical_trans}(B) shows the transition
point where the current order vanishes as seen from the crossing of $B_L$ curves \cite{binder}
for different $L$.
Remarkably, we find that loop current order persists into the regime where the superfluid
order is absent, revealing an intermediate insulating phase with
staggered loop currents - this is the CMI. 

For $J_\perp/J=1$, we find the BKT transition
occurring at $1/J_\tau=0.887(1)$ while the current order vanishes at the Ising transition
which is located at $1/J_\tau=0.981(4)$, where the error bars on the transition point are
estimated from the error in the location of the dip in the inset of Fig.\ref{Fig:classical_trans}(A)
and the error in the crossing point in Fig.\ref{Fig:classical_trans}(B), both of which yield the
limiting thermodynamic values for the transition points. This establishes that the phase
diagram supports {\it three} phases: CSF, CMI, and MI. A similar analysis for different
values of $J_\perp$ allows us to obtain the phase diagram in Fig.~\ref{Fig:classical_phased}(A).

We have already seen that the
transition out of the CSF, i.e., the CSF-CMI transition, is of the BKT type. 
The scaling
of the divergent susceptibility peak $\chi_{\rm crit}(L)$ for current order 
(Fig.~\ref{Fig:classical_trans}(B) inset) shows that the CMI-MI critical point 
is a 2D Ising transition.
Such consecutive, closely spaced, BKT-Ising thermal transitions are also observed in 
the classical 2D FFXY
model \cite{olsson.prl1995}, although its Hamiltonian is quite
distinct from $H_{\rm XY}$, and the chiral order in the classical model corresponds
to having in-plane currents rather than interlayer currents as in our bilayer model.
Such consecutive transitions are also found in
spinor condensates \cite{mukerjee.prb2009}.

\begin{figure}
\includegraphics[width=2.8in,height=1.7in]{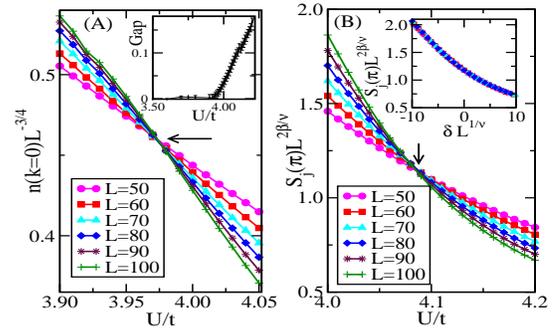}
\caption{(Color online)
(A) DMRG results for $n(k=0)L^{-3/4}$ versus $U/t$, for the FFBH Hamiltonian in Eqn.~\ref{LadderHam}
with $t_\perp\! =\! t$ and various $L$. The crossing of these curves at $U_{c1}/t \approx 3.98(1)$ 
yields the
CMI-MI transition (see text). Inset shows the onset of the charge gap at $U_{c1}$.
(B)  Rung current structure factor $S_j(\pi)L^{2\beta/\nu}$ versus $U/t$ at $t_\perp=1$.
The intersection point yields the CMI-MI Ising
transition at $U_{c2} \approx 4.08(1) t$. Inset shows $S_j(\pi)L^{2\beta/\nu}$ versus $\delta L^{1/\nu}$ with $\delta \equiv
(U - U_{c 2})/t$, for different $U/t$, leading to a scaling collapse for 2D Ising
exponents $\nu=1$ and
$\beta=1/8$.}
\label{Fig:dmrg}
\end{figure}

{\it DMRG study. ---}
We next study the FFBH ladder model in Eq.~(\ref{LadderHam}) at a filling of
one boson per site using the finite size DMRG (FS-DMRG) method \cite{whitedmrg.prl1992}.
(We set $t=1$ here.)
As noted previously \cite{sengupta.epl2011,cha.pra2011}, the boson momentum distribution
$n(k)$ in the presence of $\pi$-flux exhibits two peaks; for our gauge choice, these
peaks are located at $k=0,\pi$. In the CSF state, which is a Luttinger liquid \cite{giamarchi} on
the ladder, we have a singular momentum distribution
$n(k \!\to\! 0) \sim |k|^{-(1-K/2)}$, with $K > 0$ being an interaction
dependent Luttinger parameter \cite{supp}. Similarly, $n(k\to\pi) \sim
|k-\pi|^{-(1-K/2)}$. Let $U_{c 1}$ denote the location of the transition out of the CSF
into an insulator. If this transition is of the BKT type, as shown
from our XY model study, the exponent $K$ should
take on a {\it universal} value $K_{c}\!\! =\!\! 1/2$ at $U_{c 1}$.
A plot of $n(k\!=\!0) L^{-3/4}$ for different
$L$ should thus show a crossing point at the transition out of the CSF, as seen at
$U_{c1} \approx 3.98(1)$ in Fig.~\ref{Fig:dmrg}(A)  for $t_\perp=1$. Remarkably, 
Fig.~4(A) (inset) shows that
the charge gap also becomes nonzero for $U > U_{c1}$, coinciding with the point
where $K\!=\! 1/2$, confirming that
the CSF-to-insulator transition is a BKT transition.
This leads to the phase boundary of
the CSF state shown in Fig.~2(B).

The staggered current order parameter can be obtained from the rung-current
structure factor 
$S_j(k)=\frac{1}{L^2}\sum_{x,x'}{e^{ik(x-x')}\langle{j_x j_{x'}}\rangle}$,
with $j_x=i \left( a_x^\dagger b_x - b_x^\dagger a_x \right)$.
$S_j(k=\pi) \sim L$ indicates long range staggered current order.
Our XY model study informs us that the current order disappears at
a MI-CMI transition which is in the Ising universality class. We thus
expect $S_j(\pi)$ to obey the critical scaling form
$S_j (\pi)L^{2\beta/\nu}= f \left(\left(U-U_{c 2} \right)L^{1/\nu}\right)$,
where $U_{c 2}$ is the CMI-MI critical point, 
$f(.)$ is a universal scaling function, and $\beta = 1/8$ and
$\nu=1$ are the Ising critical exponents.
As a result, curves of $S_j (\pi)L^{2\beta/\nu}$ for different $L$ are expected to
intersect at the MI-CMI critical point $U_{c 2}$. 
This crossing, as seen at $U_{c 2} \approx 4.08(1)$ for $t_\perp=1$
from Fig.~\ref{Fig:dmrg}, allows us to carefully locate the CMI-MI phase transition.
As seen in Fig.\ref{Fig:dmrg} (inset), plotting $S_j (\pi)L^{2\beta/\nu}$ as a
function of $(U-U_{c 2}) L^{1/\nu}$
shows a complete data collapse for $U_{c2} = 4.08$. Similar to our discussion for the
computations on the XY model, our analysis of these crossing points in the FFBH model
yields the limiting thermodynamic values of the transition points, and the error bars
are estimated
from examining the errors in these crossing points.
Such an analysis, carried out for a range of values of $t_\perp/t$,
allows us to map out the MI-CMI phase boundary in Fig.~\ref{Fig:classical_phased}(B);
we find $U_{c2} > U_{c1}$, again consistent with an intermediate CMI state.

{\it Discussion. ---}
Our computations on the FFBH model at unit filling
and the XY model (which describes the FFBH model at large integer filling),
suggest that the CMI appears near the tip of the Mott lobes at
all boson fillings on the ladder. We have generalized the work of
Ref.~\cite{sorella.prl2007} to obtain a long-range Jastrow
correlated wavefunction which captures all the essential correlations
of this CMI state on the ladder \cite{supp}.
Since the CMI is {\it completely} gapped, with not just a
charge gap but also an ``Ising'' gap to charge-neutral excitations, it 
will be stable in a 2D system of weakly coupled FFBH ladders.

The CSF and CMI states are bosonic analogs of staggered current
metallic \cite{ddw} and insulating \cite{marston.prb2002} states
of fermions
in models of cuprate superconductors.
The CSF and CMI also 
find analogs in insulating magnets: paramagnetic
gapless \cite{chiralgapless} or spin-gapped \cite{chiral} phases
with long range vector chiral order.


The CMI  may be realized in
a Josephson junction ladder at a magnetic field of $hc/4e$ flux per plaquette
 \cite{mooij}, where it
 would appear as an insulator in transport measurements. With a Josephson coupling $\sim 1K$,
we estimate that the spontaneous loop currents could produce staggered magnetic fields
$\sim 1$nT for arrays with lattice parameter $10 \mu$m, which could be measured using SQUID microscopy
 \cite{fong.revsci2005}.
 Ultracold bosonic atoms in the presence of a (uniform or staggered)
synthetic $\pi$-flux \cite{spielman.nature2009}
are candidates to realize the CMI. The signature of the flux would appear as twin
peaks in the atom momentum distribution: the peaks
would be sharp in the CSF
but broad in the CMI and MI. Re-interfering
the $k=0$ and $k=\pi$ peaks obtained in time of flight via Bragg pulses
\cite{bragg.epjd2005} could test for the persistence of
intermode coherence (the phase $\theta=\pm\pi/2$)
in the CMI, and distinguish it from the MI. 
Jaynes-Cummings lattices in a ``magnetic field'' \cite{JClattice}, could also be used to simulate
a polariton FFBH model.

{\it Acknowledgments:}
We thank B. P. Das, M. P. A. Fisher, D. A. Huse, and J. H.
Thywissen, for discussions. We acknowledge support from DST, Govt. of India
(SM and RVP), CSIR (RVP), and NSERC of Canada (AP).


\begin{thebibliography}{51}
\bibitem[{foo()}]{foot1}
\bibinfo{note}{Our main results are independent of this gauge choice.}
\bibitem[{sup()}]{supp}
\bibinfo{note}{Details will be published elsewhere.}
\end{thebibliography}
\end{document}